# Disentangling Phonon Channels in Nanoscale Thermal Transport

Samik Mukherjee[1,*], Marcin Wajs[1], Maria de la Mata[2,¥], Uri Givan[4], Stephan Senz[4], Jordi Arbiol[2,3], Sebastien Francoeur[1], and Oussama Moutanabbir[1,*]

[1]*Department of Engineering Physics, École Polytechnique de Montréal, C. P. 6079, Succ. Centre-Ville, Montréal, Québec H3C 3A7, Canada*
[2]*Catalan Institute of Nanoscience and Nanotechnology (ICN2), CSIC and BIST, Campus UAB, Bellaterra, 08193 Barcelona, Catalonia, Spain*
[3]*ICREA, Pg. Lluís Companys 23, 08010 Barcelona, Catalonia, Spain*
[4]*Max Planck Institute of Microstructure Physics, Weinberg 2, Halle (Saale) 06120, Germany*
[¥]*Present Address: Departamento de Ciencia de los Materiales, Ing. Met. y Qca. Inorg., IMEYMAT, Universidad de Cádiz, 11510 Puerto Real, Spain*
[*] Email: samik.mukherjee@polymtl.ca; oussama.moutanabbir@polymtl.ca

**Abstract:**

Phonon surface scattering has been at the core of heat transport engineering in nanoscale structures and devices. Herein, we demonstrate that this phonon pathway can be the sole mechanism only below a characteristic, size-dependent temperature. Above this temperature, the lattice phonon scattering co-exist along with surface effects. By artificially controlling mass disorder and lattice dynamics at the atomic-level in nanowires without affecting morphology, crystallinity, chemical composition, and electronic properties, the *temperature-thermal conductivity-diameter* triple parameter space is mapped, and the main phonon scattering mechanisms are disentangled. This led to the identification of the critical temperature at which the effect of lattice mass-disorder on thermal conductivity is suppressed to an extent that phonon transport becomes governed entirely by the surface. This behavior is discussed based on Landauer-Dutta-Lundstrom near-equilibrium transport model. The established framework provides the necessary input to further advance the design and modelling of phonon and heat transport in semiconductor nanoscale systems.

Phonon transport in a crystal is controlled by surface as well as lattice effects[1,2]. The latter dominate in bulk materials and include isotopic-disorder, charge carrier, defect, and anharmonic scattering channels. Surface scattering is, however, prevalent in nanoscale materials and has been at the center of the recently expended efforts to engineer thermal transport at the nanoscale[3–9]. In this regard, semiconductor nanowires (NWs) have been extensively studied as a unique playground for phonon engineering that combines shape anisotropy and large surface-to-volume ratio[10–16]. In such quasi-one-dimensional nanostructures, the phonon-surface interaction renormalizes the phonon lifetime and/or the non-equilibrium phonon population distribution[3,9,17]. The interaction with the rough and amorphous surface showed evidence of Anderson localization of phonons in NWs[4,5]. It was also demonstrated that surface interaction shortens the dominant phonon wavelength and mean-free path in the NW heat spectra relative to bulk values[6]. Additionally, the surface roughness and its spatial correlation length were also found to tune the NW thermal conductivity below the Casimir limit[7–9].

Despite the expected overwhelming dominance of surface scattering, lattice effects must also persist because of the NW finite diameter. For example, calculations showed that a small degree of lattice disorder in alloyed NWs can act like a low-pass filter, suppressing the high-frequency phonon modes while leaving the low-frequency modes unaffected[18]. A separate calculation revealed a complementary effect, wherein a short-range spatial correlation in the mass-disorder of one-dimensional structures acts like a high-pass filter, showing a dramatic increase in the transmittance of high-frequency phonon modes and suppression of the low-frequency modes[19]. Thermal transport in NWs was also reported to be anisotropic, depending on the crystal orientation[20]. For NWs of same diameter, such behavior could only be attributed to the anisotropy of the acoustic phonon group velocity along different crystal orientations. The aforementioned studies indicate that along with surface scattering, thermal transport in such quasi-one-dimensional structures underlies a subtle but important contribution from the crystalline lattice as well. However, the individual contributions of these mechanisms remain hitherto undiscovered and experimentally inaccessible. This work addresses this fundamental question by establishing an experimental framework to track and distinguish the inherently intertwined lattice and surface scatterings. By tailoring the mass-disorder in silicon (Si) NWs, while keeping the chemical composition, crystallinity, electronic properties, and morphology intact, the work demonstrates that the temperature-dependent effects and the size-dependent effects on phonon transport can indeed be disentangled.

**Nanoscale engineering of lattice disorder**

Mass-disorder engineering is achieved during the NW growth in ultrahigh vacuum chemical-vapor-deposition (UHV-CVD) chamber using isotopically enriched silane precursors $^{28}SiH_4$ and $^{30}SiH_4$ with an isotopic purity higher than 99.9%[21] (see Methods for details on NW growth). The NWs discussed in this work are either isotopically pure $^{30}Si$ NWs or isotopically mixed $^{28}Si_x{}^{30}Si_{1-x}$ (x~0.4) NWs. The two sets of NWs are therefore chemically, electrically, and morphologically identical. The NWs are ~ 10 μm-long with an average diameter of ~55 nm. Fig. 1(a) shows a representative high-angle annular dark field (HAADF) image of an isotopically mixed $^{28}Si_x{}^{30}Si_{1-x}$ NW analyzed in an aberration-corrected scanning transmission electron microscope (STEM) (see Methods). The panels 2 and 3 in the figure show the HAADF/STEM images at progressively increasing magnification. The fast Fourier transform (FFT) indicates the NW growth orientation. The STEM analysis of isotopically lattice-ordered $^{30}$Si NWs (not shown here) exhibits similar microstructure.

Fig. 1(b) displays the Si-Si longitudinal optical (LO) phonon mode of single $^{30}$Si (left) and $^{28}$Si$_x$$^{30}$Si$_{1-x}$ (right) NWs, recorded at three different base temperatures. The data indicates a consistent redshift of the peak position in both sets of NWs, as the base temperature increases. Fig. 1(c) summarizes this evolution over a broader temperature range of 4-300 K. The reasons for the observed redshift lie in the anharmonic effects (phonon-phonon coupling) and the quasi-harmonic effects (changes in crystal volume), both being temperature-dependent. The frequency of a mode at a temperature T is given by $\omega(T) = \omega_0 + [\Delta\omega(T)_{vol}] + [\Delta\omega(T)_{anh}]$, where '$\omega_0$' is the frequency as T approaches 0 K, $\Delta\omega(T)_{vol}$ is the shift in frequency caused by changes in crystal volume, and $\Delta\omega(T)_{anh}$ is the frequency shift caused by the anharmonic phonon-phonon coupling. $\omega(T)$ is expressed as[22,23]

$$\omega(T) = \omega_0 + [\omega_0\{e^{-3AT} - 1\}] + \left[B\left\{1 + \frac{2}{e^y - 1}\right\} + C\left\{1 + \frac{3}{e^z - 1} + \frac{3}{(e^z - 1)^2}\right\}\right], \quad (1)$$

where $y = \hbar\omega_0/2k_BT$ and $z = \hbar\omega_0/3k_BT$. B and C are cubic and quartic anharmonic constants denoting the strength of the three- and four-phonon anharmonic coupling processes, respectively. Details related to the parameter 'A' can be found in Section 2 of the Supplementary Information. The data in Fig. 1(c) was fitted using equation (1) (solid lines) yielding $\omega_0$ of (512.3 ± 0.3) cm$^{-1}$ and (521.7 ± 0.3) cm$^{-1}$ for $^{30}$Si and $^{28}$Si$_x$$^{30}$Si$_{1-x}$ NWs, respectively. $\omega_0$ is a key parameter for the subsequent analysis. The data in Fig. 1(c) also indicate that quartic anharmonicity can be excluded from the observed temperature-induced redshift of the LO modes, since the coupling constant C was found to be negligible for both sets of NWs. The total redshift in ω from 4 to 300 K is insensitive to the isotopic content. This is expected, since neither the anharmonic effect nor the quasi-harmonic effect depend on the lattice-disorder[24].

**Local heat in mass disorder-engineered NWs**

Next, the evolution of LO mode peak positions ($\omega_{PD}$) were measured as a function of the incident laser power density, at fixed base temperatures. Examples showing the evolution of $\omega_{PD}$ as a function of incident power density at 140 K are displayed in Figs. 2(a) and (b) for $^{30}$Si NWs and $^{28}$Si$_x$$^{30}$Si$_{1-x}$ NWs, respectively. $\omega_{PD}$ typically undergoes a redshift with increasing power density, indicative of laser-induced local heating of the NWs[25,26]. At a first glance, it is clear that for the same change in the laser power-density, the total redshift in $\omega_{PD}$ of $^{28}$Si$_x$$^{30}$Si$_{1-x}$ NWs (Fig. 2(b)) is larger than that of $^{30}$Si NWs (Fig. 2(a)). This means that $^{28}$Si$_x$$^{30}$Si$_{1-x}$ NWs are heating up much faster than $^{30}$Si NWs. The effect can be best summed up by extracting the NW local temperature ($T_{NW}^{rise}$) as a function of the laser-power density. $T_{NW}^{rise}$ is estimated from equation (1), after expanding the exponential terms in a Maclaurin series and terminating at the linear term. Note that ω(T) in equation (1) is replaced with $\omega_{PD}$ and the base temperature (T) by $T_{NW}^{rise}$. $T_{NW}^{rise}$ is the sum of the base temperature (T) and the increase in the NW local temperature due to laser-induced heating. Combining all these factors, the expression of $T_{NW}^{rise}$ is obtained:

$$T_{NW}^{rise} = \frac{(\omega_0 + B) - \omega_{PD}}{\left[3A\omega_0 - \frac{4k_B B}{\hbar\omega_0}\right]} \quad (2)$$



Note that $\omega_0$ as well as the parameters A and B in the right-hand-side of equation (2) are known for both NW types. The evolution of $T_{NW}^{rise}$ as a function of the laser power density for $^{30}$Si NWs and $^{28}$Si$_x$$^{30}$Si$_{1-x}$ NWs is also displayed in Figs. 2(a) and (b), respectively. The rate of change in $T_{NW}^{rise}$ with incident power density is inversely related to $\kappa_T$ (thermal conductivity)[21,27]. Higher the $\kappa_T$, easier it is for the laser-induced heat to flow away and slower is the rate of increase in $T_{NW}^{rise}$ with incident power density. Note that only the linear regime of $T_{NW}^{rise}$ vs. power density graph has been used in the subsequent analysis. Fig. 2(c) compares this linear regime of the $T_{NW}^{rise}$ vs. power density graph for $^{30}$Si NWs and $^{28}$Si$_x$$^{30}$Si$_{1-x}$ NWs at 140 K. The ratio of the respective slopes shows that $T_{NW}^{rise}$ of $^{28}$Si$_x$$^{30}$Si$_{1-x}$ NWs increases $\sim$1.30× faster with power density than that of $^{30}$Si NWs. This indicates that at 140 K, $\kappa_T$ of $^{28}$Si$_x$$^{30}$Si$_{1-x}$ NWs is $\sim$30 % lower than that of $^{30}$Si NWs. This reduction in $\kappa_T$ results from the isotope scattering in lattice-disordered $^{28}$Si$_x$$^{30}$Si$_{1-x}$ NWs. Fig. 2(d) shows the evolution of $T_{NW}^{rise}$ at 60 K. Note that the ratio of the slopes reaches unity which indicates similar $\kappa_T$ for both isotopically mixed and isotopically pure NWs. This means that the effect of isotope-induced mass-disorder has been suppressed at 60 K, leaving surface scattering as the dominant mechanism determining the overall phonon lifetime. Also, note the difference in the slope of $T_{NW}^{rise}$ at 140 K and that at 60 K. For example, for $^{30}$Si NWs at 60 K the slope is $\sim$1.45× larger than that at 140 K, indicating that $\kappa_T$ at 60 K is reduced by $\sim$45 % w.r.t. that at 140 K. The temperature evolution of $\kappa_T$ of both the NW sets is discussed next.

Fig. 3(a) exhibits the evolution of $\kappa_T$ with temperature of both NW sets. In this work, $\kappa_T$ is always given relative to that of $^{30}$Si NWs at 300 K. The temperature evolution of $\kappa_T$ is quite different from bulk materials[28–30]. At low temperatures, when the phonon mean-free path is long and comparable to the sample dimensions, surface scattering (independent of temperature) is dominant and $\kappa_T$ of a bulk semiconductor essentially follows the $\sim T^3$ dependence of the phononic specific heat[31]. At higher temperatures, anharmonic phonon-phonon scattering starts to dominate the phonon lifetime and $\kappa_T$ scales as $\sim T^{-1}$. A bulk sample reaches the maximum value of $\kappa_T$ ($\kappa_T^{max}$) between these two temperature regimes. $T_{max}$, the temperature at which a bulk sample reaches $\kappa_T^{max}$ is material specific, varying from $\sim$16.5 K for Ge[29], to $\sim$26 K for Si[28], to $\sim$75 K for diamond[32]. It is around $T_{max}$ that the isotope effect on $\kappa_T$ of bulk semiconductors is the most prominent. This difference in the evolution of $\kappa_T$ with temperature in NWs is related to phonon surface scattering that competes with all other scattering mechanisms, even at elevated temperatures[33–35]. In the following, this behavior is further discussed based on the Landauer-Dutta-Lundstrom (LDL) transport model[36,37]. According to this model, the lattice thermal conductance ($K_T$) under near-equilibrium conditions is given by[37,38]

$$K_T = \sum_{j=LA,TA} \frac{k_B^2 T}{h} \int_j M_j^{ph}(\omega) W^{ph}(\omega, T) \mathcal{T}_j^{ph}(\omega, T) d\omega, \quad (3)$$

where the summation is over the longitudinal acoustic (LA) and the doubly degenerate transverse acoustic (TA) phonon modes. $M_j^{ph}(\omega)$ is the total number of channels available to the mode 'j' for heat conduction at an energy of $\hbar\omega$ and is given by the cross-section (S) of the conductor times the phononic density of states, $D_j^{ph}(\omega)$. $W^{ph}(\omega, T)$ is the energy- and temperature-dependent window function that selects only those channels, out of all the available channels, that can take part in thermal conduction and is given by $W^{ph}(\omega, T) = (\hbar\omega/k_B T)^2(-\partial n_{Bose}/\partial \omega)$, where $n_{Bose}$ is the equilibrium Bose-Einstein factor given by $n_{Bose} = [e^{\hbar\omega/k_B T} - 1]^{-1}$. $\mathcal{T}_j^{ph}(\omega, T)$ is the transmittance function which defines the probability that the j$^{th}$ mode is transmitted along the length (L) of the heat conductor, without backscattering. $\mathcal{T}_j^{ph}(\omega, T)$ can be approximated as $\lambda_j(\omega, T)/L$, where $\lambda_j(\omega, T)$ is the backscattering mean-free path. $\lambda_j(\omega, T)$ can in turn be expressed as $f v_j \tau_j(\omega, T)$, where $v_j$ is velocity of the mode 'j' along the direction of transmission, $\tau_j(\omega, T)$ is the mode lifetime at an energy $\hbar\omega$ and temperature T, and $f$ is a pre-factor which accounts for the backscattering mean-free path.

**Mapping the temperature-thermal conductivity-diameter space**

Various scattering mechanisms were considered in order to calculate the total phonon lifetime (see details in Section 3 of the Supplementary Information). This includes the normal anharmonic scattering of phonons[39], the Umklapp anharmonic scattering[34], scattering of phonons from isotopic mass-disorder[40], and the surface phonon scattering[29,41]. Additionally, the decay of high-energy optical phonons into acoustic phonons was also considered. The scattering of optical phonons is usually neglected due to the flatness of their dispersion curves in Si, making them rather ineffective heat carriers. However, earlier works on theoretical formalism of thermal transport in Si NWs argued that the decay of optical phonons into acoustic phonons at an energy $\hbar\omega$ can be considered as a generation rate of acoustic phonons, which partially counteracts their scattering rate at the same energy[42,43]. Finally, the total acoustic phonon scattering rate was calculated according to the Matthiessen rule. Phonon confinement is neglected due to the relatively large diameter of NWs. To this end, a Debye-like phonon spectrum was considered, wherein $D_j^{ph}(\omega) = \omega^2/2\pi^2 v_j^3$ shows a parabolic dependence on $\omega$[44]. The experimental data in Fig. 3(a) was fitted using equation (3), with the specularity factor (P) and the constants $\mathcal{C}_{LA(TA)}^{M(P)}$ (see Section 3 of the Supplementary Information for details) as free parameters. The upper limit of the integral in equation (3) was initially set at Debye frequencies $[\omega_{D,LA(TA)} = k_B\theta_{D,LA(TA)}/\hbar]$ in Si: 76.7 THz for the LA mode and 31.4 THz for the TA mode. The validity of this approach was also confirmed by comparing simulated $\kappa_T$ vs. T of bulk $^{Nat}$Si and isotopically pure bulk $^{30}$Si to experimental data[28,45]. The results are shown and discussed in Section 4 of the Supplementary Information. The parameters extracted from the simulated data of the bulk materials match the experimental values[28,45] with 5% uncertainty, justifying the approximations and with it the accuracy of the LDL model in predicting temperature evolution of $\kappa_T$.

Next, the analysis was carried out on the data for $^{30}$Si NWs in Fig. 3(a). Clearly, the model (solid black line) disagrees with the experimental data, more so above 50 K. The reasons for this discrepancy lie in the bulk-like Debye phonon spectrum, which might overestimate the contribution of phonon Umklapp scattering rate[41] and that of $M_j^{ph}(\omega)$[46] in equation (3). The optimization was done by choosing a lower cut-off frequency on a trial-and-error basis[41]. The solid green line in Fig. 3(a) was obtained after the cut-off optimization, which yielded a cut-off frequency $(\omega_{C,LA})$ of 51.7 THz for the LA mode. The optimization also found that lowering the cut-off frequency of the TA mode impacts the quality of the fit only nominally (see Section 5 of the Supplementary Information). The upper limit of the integral for the TA mode in equation (3) was therefore maintained at its Debye frequency of 31.4 THz all throughout, to avoid an additional undetermined parameter in the model. The values of P and $\mathcal{C}_{LA(TA)}^P$ for $^{30}$Si NWs were found to be 0.26 and 14.86 × 10$^4$ s$^2$ (22.01 × 10$^6$ s$^4$), respectively. For $^{28}$Si$_x$$^{30}$Si$_{1-x}$ NWs, the value of P was fixed to the same value as that of $^{30}$Si NWs, while only $\mathcal{C}_{LA(TA)}^M$ were kept as free parameters. The two sets of NWs were grown under identical conditions and mass-disorder is the only difference between them. Consequently, there is no reason that the specularity factor of surface phonon



scattering will be any different between them. The fit to the experimental data of $^{28}Si_x{}^{30}Si_{1-x}$ NWs is shown by the solid red line in Fig. 3(a), which gives $\mathcal{C}_{LA(TA)}^{M}= 17.34 \times 10^4 \, s^2$ ($20.21 \times 10^6 \, s^4$). With all parameters optimized, equation (3) was then used to evaluate $\kappa_T$ by varying the diameter ($d_{NW}$) from 25 nm up to 250 nm. The results are shown in Fig. 3(b) at the two extreme ends of the range ($d_{NW}$ of 25 nm and 250 nm). See examples of the simulated $\kappa_T$ vs. T data for NWs at a few intermediate values of $d_{NW}$ in Section 6 of the Supplementary Information. The temperature behavior of $\kappa_T$ in Fig. 3(b) is remarkably sensitive to $d_{NW}$, which can be attributed to the diameter dependence of the contribution of surface scattering to the phonon lifetime in the NWs.

From the simulated data, four parameters related to $\kappa_T$ of the NWs are evaluated next, all of which were found to depend strongly on $d_{NW}$. Of these four, one key parameters that is crucial here is the temperature limit of mass disorder-induced phonon scattering $\left(T_L^{d_{NW}}\right)$, or the temperature at which any effect of the lattice on $\kappa_T$ vanishes due to the overwhelming influence of phonon surface scattering. Fig. 4(a) shows that the evolution of $T_L^{d_{NW}}$ as a function of $d_{NW}$. The dependence on $d_{NW}$ is very close to being hyperbolic, as extracted from the power-law fit. The hyperbolic dependence is also confirmed by the log-log plot displayed in the inset of Fig. 4(a), which shows a slope of $-1.1$. Given that $T_L^{d_{NW}}$ is the temperature below which lattice mass-disorder effect on $\kappa_T$ is totally suppressed, it is also interesting to investigate how it varies as a function of the mass-variance. To this end, $\kappa_T$ vs. T data of $^{28}Si_x{}^{30}Si_{1-x}$ NWs was also simulated by varying the fractional composition of $^{28}Si$ (see an example in Section 6 of the Supplementary Information) and the dependence of $T_L^{d_{NW}}$ on the $^{28}Si$ concentration is shown in Fig. 4(b), indicating a linear behavior wherein $T_L^{d_{NW}}$ decreases by $\sim 0.85$ K for every 10.0 at.% increment



in the $^{28}$Si concentration[47]. Interestingly, the influence of the mass-variance on $T_L^{d_{NW}}$ is fairly independent of $d_{NW}$, as reflected by the same slope of the NWs at the diameter extremities of 25 nm and 250 nm. Shown in Section 7 of the Supplementary Information are the other three parameters which include: the reduction in room temperature $\kappa_T$ of $^{28}Si_x{}^{30}Si_{1-x}$ NWs relative to $^{30}$Si NWs $\left(R_{RT}^{d_{NW}}\right)$; the temperature at which $\kappa_T$ reach the maximum value $\left(T_{max}^{d_{NW}}\right)$; and the reduction in $\kappa_{max}$ of $^{28}Si_x{}^{30}Si_{1-x}$ NWs relative to $^{30}$Si NWs at $T_{max}^{d_{NW}}$ $\left(R_{T_{max}}^{d_{NW}}\right)$. $T_{max}^{d_{NW}}$ was found to have a $\sim(d_{NW})^{-1.4}$ dependence on $d_{NW}$, while $R_{RT}^{d_{NW}}$ and $R_{T_{max}}^{d_{NW}}$ were found to depend on $d_{NW}$ as $\sim\exp(-d_{NW})^{0.65}$ and $\sim\exp(-d_{NW})^{0.72}$, respectively.

**Conclusion**

The analyses above outline a framework to evaluate the *temperature-$\kappa_T$-diameter* triple-parameter space by artificially controlling mass-disorder and lattice dynamics in NWs. The work also demonstrates that phonon transport in NWs is indeed not all about the surface and there exists a temperature regime where the lattice starts contributing to the heat transport in a nanoscale structure. The framework identifies the boundary between this regime and the purely surface-dominated one and elucidates how this boundary evolves as a function of the NW diameter and the extent of mass-disorder in the crystalline lattice. By unravelling these heretofore unexplored fundamental aspects, this work disentangles the primary phonon scattering channels that strongly influences the thermal behavior at the nanoscale. This knowledge is crucial to guide the design and modelling of future nanoscale phononic and thermoelectric devices and heat management platforms.




## ACKNOWLEDGEMENTS

O.M. acknowledges support from NSERC Canada (Discovery, SPG, and CRD Grants), Canada Research Chairs, Canada Foundation for Innovation, Mitacs, PRIMA Québec, and Defence Canada (Innovation for Defence Excellence and Security, IDEaS). ICN2 acknowledges funding from Generalitat de Catalunya 2017 SGR 327. ICN2 is supported by the Severo Ochoa program from Spanish MINECO (Grant No. SEV-2017-0706) and is funded by the CERCA Programme / Generalitat de Catalunya. The HAADF-STEM microscopy was conducted in the Laboratorio de Microscopias Avanzadas at Instituto de Nanociencia de Aragon-Universidad de Zaragoza. Authors acknowledge the LMA-INA for offering access to their instruments and expertise.


## AUTHORS CONTRIBUTION STATEMENT

S.M. analyzed the data and carried out all systematic experimental and theoretical studies. M.W., supervised by S.F., helped in the low-temperature Raman studies. M.d.l.M. performed TEM analysis under the supervision of J.A.. U.G. and S.S. were involved in the initial nanowire growth work. O.M. conceived the project. S.M. and O. M. wrote the manuscript and all authors comment it.

## COMPETING INTERESTS

The authors declare no competing financial interests.

**Methods**

i) **Growth of isotopically engineered NWs:** $^{Nat}SiF_4$ gas was first enriched into isotopically enriched $^jSiF_4$ (j= 28, 29, and 30) inside a Zippe-centrifuge. This was followed by reduction of the isotopically enriched $^jSiF_4$ gases into the $^jSiH_4$ precursors, following the chemical reaction: $^jSiF_4 + 2CaH_2 \rightarrow {}^jSiH_4 + CaF_2$. The vapor-liquid-solid (VLS) method was implemented to grow the NWs, with gold (Au) acting as the catalyst. For this, Si substrates underwent degreasing in acetone, followed by a dip in 2% hydrofluoric (HF) acid for 2 mins to remove the native oxide, followed by rinsing in de-ionized water. Next, 2 nm thick Au films were then deposited on the substrates before loading them into the UHV-CVD chamber. Au-Si eutectic droplets were formed by annealing the substrate at 450°C for 20 min. The NWs started to grow with the injection of the enriched $^jSiH_4$ precursors, at a flow rate of 2 sccm. The growth temperature and the chamber pressure were maintained 480 °C and 1.5 mbar, respectively.

ii) **Scanning transmission electron microscopy:** The HAADF-STEM images of the NWs were recorded in a probe-corrected FEI Titan 60-300, equipped with a high brightness field emission gun and a CETCOR corrector from CEOS.

iii) **Raman Spectroscopy:** The Raman measurements were performed in a custom micro-Raman setup at $\lambda_{ex} = 532$ nm laser excitations. The laser was focused on the sample with a 100X (NA 0.85) objective with a spatial resolution approaching the diffraction limit. The laser power density incident on the sample was measured using a hand-held power meter. The spectra were acquired using a liquid-nitrogen cooled charged-coupled device camera (JY Symphony) mounted on a Jobin-Yvon Triax iHR550 spectrometer (grating 1,800 g·mm$^{-1}$) with a precision of 0.2 cm$^{-1}$. During the laser exposure, the sample was mounted inside a liquid-He cryostat that enabled adjusting the base temperature from 300 K down



to 4 K, with a temperature stability of 80 mK. The samples were held under high vacuum of below 10$^{-6}$ Torr. The spectrometer calibration was done using the Raman peak position of bulk silicon at 520.8 cm$^{-1}$. Prior to the Raman measurements, a suspension was first prepared by putting a small piece of as-grown sample in acetone under ultrasonic vibration for 10 min. The suspension was then dispersed on two different type of substrates: first, an Au-coated Si substrate and second, a commercial Au-grid with 4 µm sized holes. The first substrate was used for investigating such intrinsic properties of the NWs (the ω *vs.* T analysis in Fig. 1(c)), wherein any laser-induced heating effect is undesirable. The Au-coated Si substrate underneath the NWs provide a good thermal contact and act as a heat sink that can quickly dissipate any laser-induced heat. The second substrate was used for the transport measurements within the framework of the power dependent study (the ω$_{PD}$ *vs.* P.D. analysis in Fig. 2). This pronounced the laser-induced heating effect of the NWs and suppressed any underlying effect of the substrate on the transport measurements.



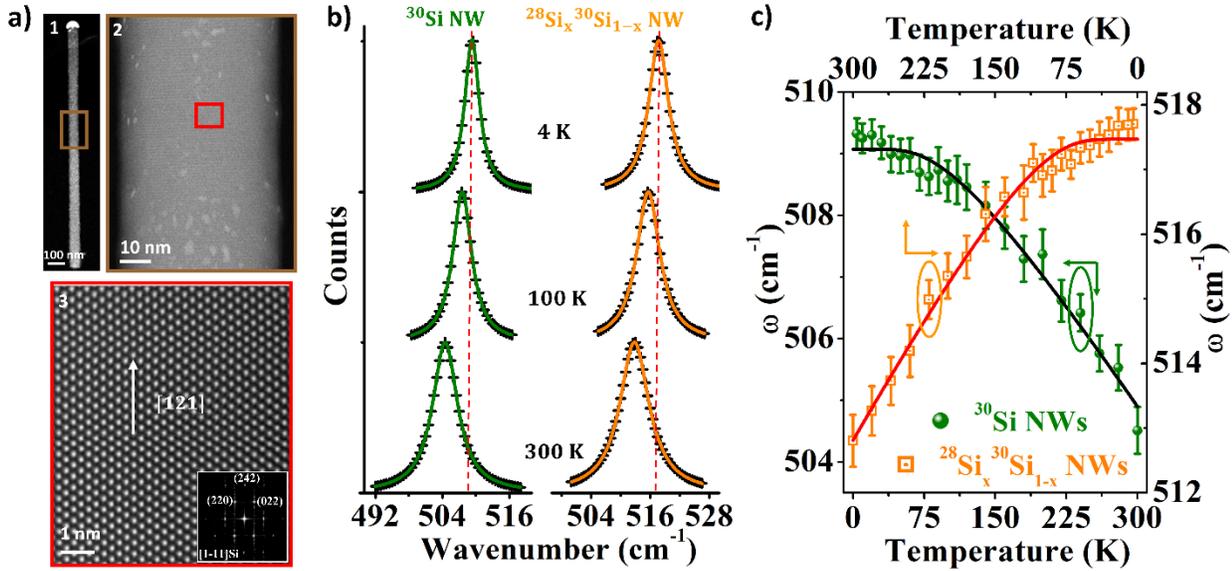

**Fig. 1 | HAADF image and Raman spectra of the isotopically programmed NWs.** (a) HAADF image from HR-STEM analysis of a $^{28}Si_x{}^{30}Si_{1-x}$ NW at different magnifications. (b) Back-scattering micro-Raman spectra, collected at different ambient temperatures, showing the LO phonon mode of a $^{30}Si$ NWs (left) and a $^{28}Si_x{}^{30}Si_{1-x}$ NWs (right). The Voigt fits are shown as solid lines. The vertical red dotted line marks the peak positions of the LO modes at 4 K. See section 1 of the supplementary information for details on laser power optimization. (c) The evolution of Si-Si peak position (ω) as a function of ambient temperature. The fit to the ω *vs.* T data is done using equation (1) (solid lines). From the fits, the parameters A in equation (1) was found to be $2.63 \times 10^{-3}$ K$^{-1}$ and $2.70 \times 10^{-3}$ K$^{-1}$, while the parameter B was found to be $-3.96$ cm$^{-1}$ and $-3.91$ cm$^{-1}$, for the $^{30}Si$ NWs and the $^{28}Si_x{}^{30}Si_{1-x}$ NWs, respectively.



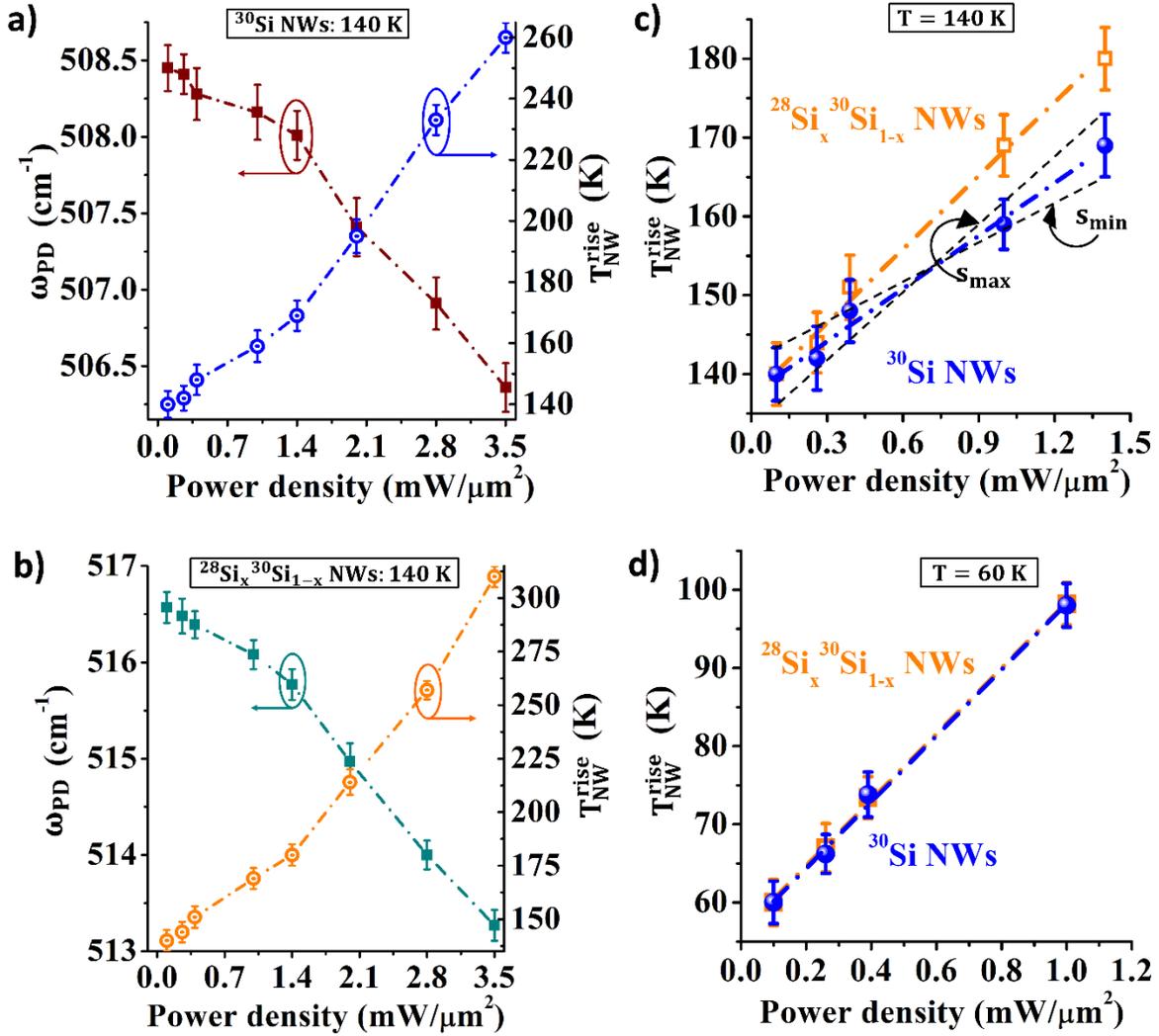

**Fig. 2 | Evolution of Raman peak position and the local temperature of the NWs as a function of laser power density** (a) Evolution of $\omega_{PD}$ and $T_{NW}^{rise}$ of $^{30}$Si NWs as a function of laser power density at 140 K; (b) Evolution of $\omega_{PD}$ and $T_{NW}^{rise}$ of the $^{28}Si_x{}^{30}Si_{1-x}$ NWs, as a function of laser power density, at an ambient temperature of 140 K. The data for $\omega_{PD}$ displayed in (a) & (b) were averaged over 10-15 individual NWs. The corresponding error bars are the standard deviation of the data. The error bars associated with the $T_{NW}^{rise}$ comes from two sources: the standard deviation (error bars) associated with the mean value of $\omega_{PD}$ and the uncertainty associated with the value of $\omega_0$ estimated from the fit; (c) Evolution in the linear regime of $T_{NW}^{rise}$ as a function of laser power density of $^{30}$Si NWs and the $^{28}Si_x{}^{30}Si_{1-x}$ NWs at 140 K. The lines indicate the linear fit to the corresponding data. The data for the $^{30}$Si NW has also been used to demonstrate how the uncertainty associated with the slope of linear fit to the $T_{NW}^{rise}$ vs. power density data was calculated. The errors bars were used to estimate the maximum possible slope ($s_{max}$) and the minimum possible slope ($s_{min}$), associated with the $T_{NW}^{rise}$ vs. power density data. The uncertainty associated



with the slope was calculated as $\Delta s = (s_{max} - s_{min})/2$; (d) Evolution of $T_{NW}^{rise}$ as a function of laser power density of the $^{30}$Si NW and the $^{28}$Si$_x$$^{30}$Si$_{1-x}$ NWs at 60 K.

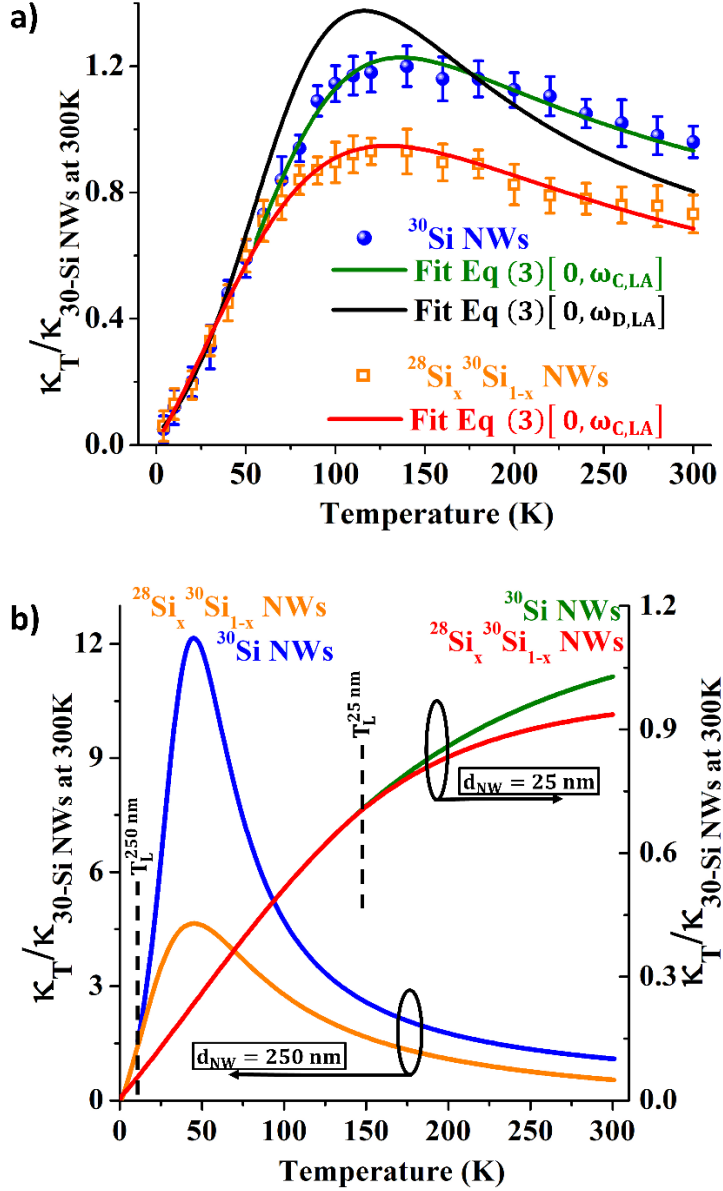

**Fig. 3 | Experimenal and simulated $\kappa_T$ of the NWs** (a) Evolution of $\kappa_T$ of $^{30}$Si NWs and $^{28}$Si$_x^{30}$Si$_{1-x}$ NWs as a function of the ambient temperature. The data are normalized to $\kappa_T$ of $^{30}$Si NWs at 300 K. The solid black line is the fit to the data of the $^{30}$Si NWs using equation (3) with $[0, \omega_{D,LA}]$ as the limits of the integration. The solid blue line is a fit to the data of the $^{30}$Si NWs using equation (3) with $[0, \omega_{C,LA}]$ as the integration boundaries. The solid red line is a fit to the data of the $^{28}$Si$_x^{30}$Si$_{1-x}$ NWs using equation (3) with $[0, \omega_{C,LA}]$ as the integration boundaries; (b) Simulated $\kappa_T$ of $^{30}$Si NWs and $^{28}$Si$_x^{30}$Si$_{1-x}$ NWs as a function of ambient temperature. The figure displays two sets of simulations corresponding to two different diameters of 250 nm and 25 nm. In order to be consistent with the experimental data in (a), the simulated data in (b) was also normalized to $\kappa_T$ of $^{30}$Si NWs at 300 K.

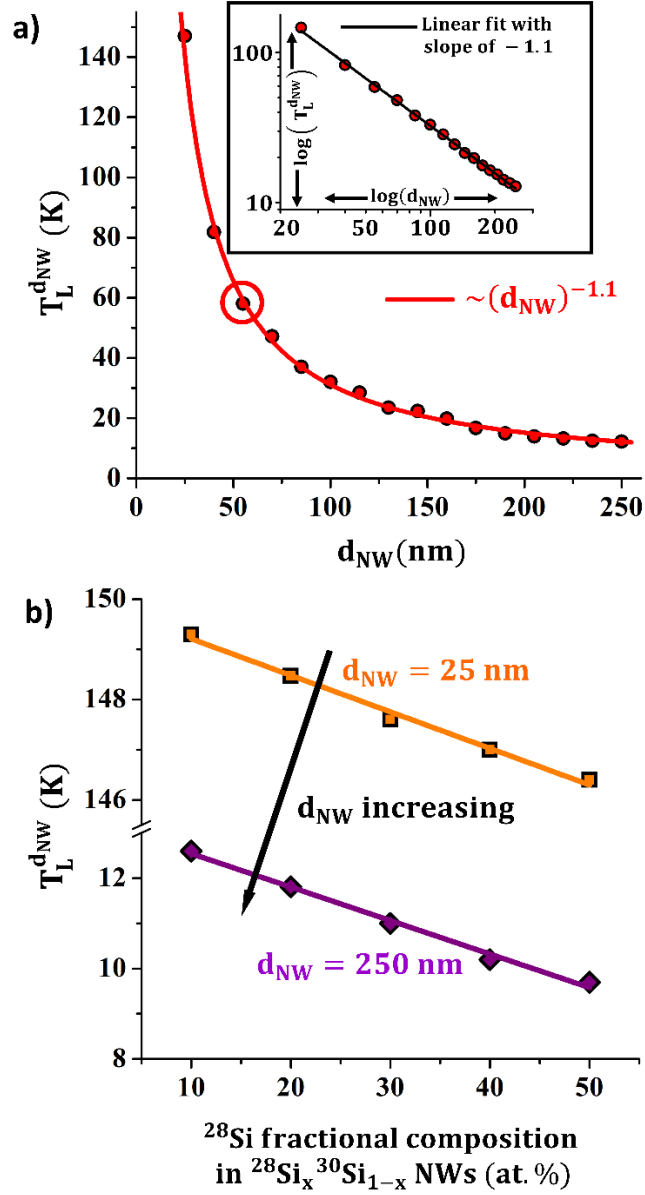

**Fig. 4 | Evolution of $T_L^{d_{NW}}$ as a function of NW diameter and mass-variance** (a) Evolution of $T_L^{d_{NW}}$ as a function of the NW diameter ($d_{NW}$). The solid red line is a power-law fit to the simulated data. Inset: Log-log plot of the $T_L^{d_{NW}}$ vs. $d_{NW}$ data. The data has been fitted with a straight line which shows a slope of $-1.1$. The simulated data point corresponding to $d_{NW} = 55$ nm (the average diameter of the NWs investigated experimentally in this work) is highlighted using the colored circle; (b) Evolution of $T_L^{d_{NW}}$ as a function of the fractional composition of $^{28}Si$ isotope within a $^{28}Si_x{}^{30}Si_{1-x}$ NW. The fractional composition is indicative of the mass-variance, the maximum mass variance being at 50.0 at.% composition of the two isotopes. The simulated data is shown only for the thinnest and the thickest NWs with diameters of 25 nm and 250 nm, respectively. The solid lines are linear fit to the respective data.

**SUPPLEMENTARY INFORMATION**

**Disentangling Phonon Channels in Nanoscale Thermal Transport**

Samik Mukherjee[1,*], Marcin Wajs[1], Maria de la Mata[2,¥], Uri Givan[4], Stephan Senz[4], Jordi Arbiol[2,3], Sebastien Francoeur[1], and Oussama Moutanabbir[1,*]

[1]*Department of Engineering Physics, École Polytechnique de Montréal, C. P. 6079, Succ. Centre-Ville, Montréal, Québec H3C 3A7, Canada*
[2]*Catalan Institute of Nanoscience and Nanotechnology (ICN2), CSIC and BIST, Campus UAB, Bellaterra, 08193 Barcelona, Catalonia, Spain*
[3]*ICREA, Pg. Lluís Companys 23, 08010 Barcelona, Catalonia, Spain*
[4]*Max Planck Institute of Microstructure Physics, Weinberg 2, Halle (Saale) 06120, Germany*
[¥]*Present Address: Departamento de Ciencia de los Materiales, Ing. Met. y Qca. Inorg., IMEYMAT, Universidad de Cádiz, 11510 Puerto Real, Spain*
[*] Email: samik.mukherjee@polymtl.ca; oussama.moutanabbir@polymtl.ca

**1) Optimization of laser power density:** Before Raman studies, a suspension was first prepared by putting a small piece of as-grown sample in acetone under ultrasonic vibration for 10 min. The suspension was then dispersed on two different type of substrates, as shown in Fig. S1(a):left. During the ($\omega$, $\Gamma$) *vs*. T measurements of the $^{30}$Si NWs and the $^{28}$Si$_x$$^{30}$Si$_{1-x}$ NWs (Fig. 1(c) of the main manuscript), the incident laser power density was optimized in the following manner to ensure the minimum laser induced heating of the NWs. The optimization was performed at 4 K. The reason being that the $\kappa_T$ of NWs is expected to be the minimum[1], and therefore the NWs have a much higher sensitivity to laser-induced heating effect at 4 K. So, if the laser power density is optimized at 4 K, it can be assumed to stay optimized for T > 4 K, all the way up to 300 K. Fig. S1(b) shows the Raman peak position as a function of power density for 3 different $^{28}$Si$_x$$^{30}$Si$_{1-x}$ NWs, at an ambient temperature of 4 K. As can be seen in Fig. S1(b), the measured peak position of the NWs shifts at power densities above 0.32 mW/µm$^2$. This blueshift is due to the laser-induced heating. The peak position stays essentially the same for power densities below 0.32 mW/μm$^2$, meaning the laser power density is low enough to induce any additional heating. The optimal power density for $^{30}$Si NWs was identified using the same procedure and found to be the same (0.32 mW/μm$^2$). The (ω, Γ) *vs.* T measurements shown in Fig. 1(c) of the main manuscript were carried out using this optimal power density of 0.32 mW/μm$^2$. The power optimization for the peak position *vs.* power density plots (Fig. 2 of the main manuscript) was performed in a similar manner, with NWs suspended atop holes on Au-grid (commercially purchased) as shown in Fig. S1(a): right.

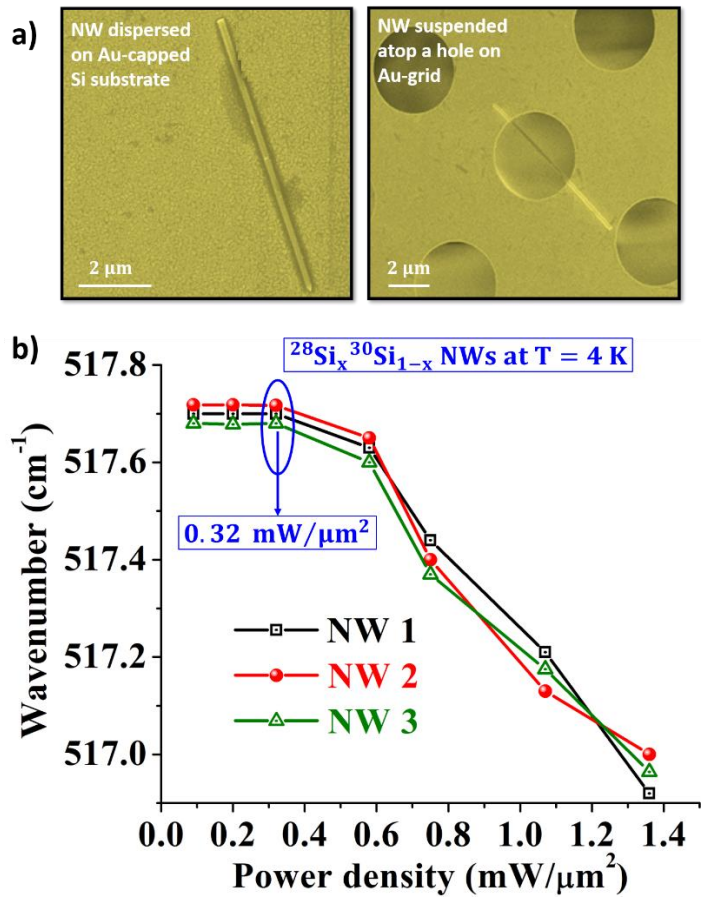

**Fig. S1 │ Sample preparation for Raman measurements and laser power optimization.** (a) Left: Scanning electron microscope (SEM) image of a single $^{28}$Si$_x$$^{30}$Si$_{1-x}$ NW, dispersed on a Au-



coated Si substrate. While investigating such intrinsic properties of the NWs (like the ω vs. T analysis in Fig. 1(c) of the main manuscript), any laser-induced heating effect is undesirable. The Au-coated Si substrate underneath the NWs provide a good thermal contact and act as a heat sink that can quickly dissipate any laser-induced heat. Right: SEM image of a single $^{28}Si_x\,^{30}Si_{1-x}$ NW, dispersed atop a hole on an Au-grid. This pronounced the laser-induced heating effect and suppressed any underlying effect of the substrate. (b) Measured Raman peak position as a function of laser power density of 3 individual $^{28}Si_x\,^{30}Si_{1-x}$ NWs, at an ambient temperature of 4 K.

**2) Details related to the expression of $\Delta\omega(T)_{vol}$:** $\Delta\omega(T)_{vol}$ is the shift in frequency caused by changes in crystal volume given by[2,3]

$$\Delta\omega(T)_{vol} = \omega_0 \left[\exp\left\{\int_0^T -n\gamma_j(T')\alpha(T')\,dT'\right\} - 1\right], \quad (E1),$$

where $\gamma_j$ is the mode-specific Grüneisen parameter, the degeneracy factor ($n$) is 3 for the LO mode in Si, and $\alpha$ is the linear thermal expansion coefficient. Both $\gamma_j$ and $\alpha$ evolve non-linearly with temperature[4,5], making the product of $\gamma_j$ and $\alpha$ rather complicated to describe using any phenomenological analytical expression. Hence, for practical purposes, mean isothermal values of $\langle\gamma\rangle_T$ and $\langle\alpha\rangle_T$ are often considered in literature for fitting the temperature evolution of phonon energies[6,7]. Here, the product of $\langle\gamma\rangle_T$ and $\langle\alpha\rangle_T$ has been considered as a single fitting parameter, A, which can be taken outside the integral in equation (E1). Equation (E1) was therefore simplified to

$$\Delta\omega(T)_{vol} = \omega_0[e^{-3AT} - 1] \quad (E2)$$



## 3) Description of the phonon scattering mechanisms:

Table 1: Summary of the scattering processes considered for calculating $\tau_j(\omega, T)$

| Scattering mechanism | Phonon mode | Scattering rates (s$^{-1}$) | Description of parameters | Reference |
|---|---|---|---|---|
| Normal anharmonic | LA | $\tau_{N,LA}^{-1}(\omega, T) = B_{N,LA}\omega^2 T^3$ | $B_{N,LA} = k_B^3 \gamma_{LA}^2 V / \bar{m}\hbar^2 v_{LA}^5$ | Ref [8,9] |
| | TA | $\tau_{N,TA}^{-1}(\omega, T) = B_{N,TA}\omega T^4$ | $B_{N,LA} = k_B^4 \gamma_{TA}^2 V / \bar{m}\hbar^3 v_{TA}^2$ | |
| Umklapp anharmonic | LA(TA) | $\tau_{U,LA(TA)}^{-1}(\omega, T) = B_{U,LA(TA)}\omega^2 T e^{-\theta_{D,LA(TA)}/3T}$ | $B_{U,LA(TA)} = \hbar\gamma_{LA(TA)}^2 / \bar{m} v_{LA(TA)}^2 \theta_{D,LA(TA)}$ | Ref [9,10] |
| Isotopic mass-disorder | LA(TA) | $\tau_{I,LA(TA)}^{-1}(\omega) = A_{LA(TA)}\omega^4$ | $A_{LA(TA)} = V g_2 / 4\pi v_{LA(TA)}^3$ | Ref [11] |
| Surface | LA(TA) | $\tau_{S,LA(TA)}^{-1} = v_{LA(TA)}/\mathcal{L}_{eff}$ | $\mathcal{L}_{eff} = [(1-P)/(1+P)W_s + (1/L)]^{-1}$ | Ref [12–14] |
| Acoustic phonon generation rate due to anharmonic scattering of optical phonons | LA(TA) | $\tau_{G_{LA(TA)}^M}^{-1}(\omega, T) = \mathcal{C}_{LA(TA)}^M \omega^3 g_2 [\coth(\hbar\omega/2k_B T)]^{1/2}$ | $\mathcal{C}_{LA(TA)}^M$ is the proportionality constant for the isotopically mixed NWs | |
| | | $\tau_{G_{LA(TA)}^P}^{-1}(\omega, T) = \mathcal{C}_{LA(TA)}^P \omega^5 [\coth(\hbar\omega/2k_B T)]$ | $\mathcal{C}_{LA(TA)}^P$ is the proportionality constant for the isotopically pure NWs | Ref [9] |

$\gamma_{LA(TA)}$ is the Grüneisen parameter for the LA(TA) mode which were taken to be 1.04 (0.56)[8]; $v_{LA(TA)}$ is the velocity of the LA(TA) mode which were taken to be 8430 m/s (5840 m/s)[8]; $\bar{m}$ is the average isotopic mass which is 29.973 amu. for the $^{30}$Si NWs and 29.015 amu. for the $^{28}Si_x{}^{30}Si_{1-x}$ NWs; V is the volume per atom which was taken to be $2.0 \times 10^{-23}$ cm$^3$ for Si[15]; $\theta_{D,LA(TA)}$ is the Debye temperature of the LA(TA) mode which were taken to be 586 K (240 K)[8]; $g_2 = \{\sum_i(c_i m_i)^2 - (\sum_i c_i m_i)^2\}/(\sum_i c_i m_i)^2$ is the second-order moment of mass-fluctuation. $c_i$ and $m_i$ are the fractional composition and the atomic mass of the $i^{th}$ isotope; P is the specularity factor for phonon surface scattering; $W_s$ and L are the physical width and length of the conductor, respectively. $W_s$ in the expression of $\tau_{S,LA(TA)}^{-1}$ was set equal to the average NW diameter of 55 nm and the contribution of L was neglected due to high NW aspect-ratio. $\hbar$ and $k_B$ have their usual meaning.

**Scattering mechanisms:** The first scattering mechanism highlighted in Table 1 is the anharmonic phonon-phonon scattering mechanism, which shows a frequency and temperature dependence of the form $\tau_{N,LA}^{-1}(\omega, T) = B_{N,LA}\omega^2 T^3$ for the LA mode and $\tau_{N,TA}^{-1}(\omega, T) = B_{N,TA}\omega T^4$ for the TA mode[8,9]. The constants $B_{N,LA}$ and $B_{N,TA}$ can be expressed as $k_B^3 \gamma_{LA}^2 V / \bar{m}\hbar^2 v_{LA}^5$ and $k_B^4 \gamma_T^2 V / \bar{m}\hbar^3 v_{TA}^2$, respectively[9]. Note, that the normal anharmonic scattering of the acoustic phonons is not a resistive process. Hence, combining this effect with the other scattering



mechanisms while fitting the data for isotopically enriched and disordered bulk Ge[13], called for an additional corrective term in the expression for $\kappa_T$. However, the contribution of this corrective term to the total $\kappa_T$ was found to be below 1% at all temperatures. Even in the Callaway's original calculations, the corrective term was initially included after having taken the normal anharmonic scattering into account, only to be subsequently neglected once its contribution was found to be insignificant[16,17]. Consequently, the corrective term has been neglected here. Second, the Umklapp scattering of phonons whose scattering rate can be shown to be[9,10] $\tau_{U,LA(TA)}^{-1}(\omega, T) = B_{U,LA(TA)} \omega^2 T \exp(-\theta_{D,LA(TA)}/3T)$, where $B_{U,LA(TA)} = \hbar \gamma_{LA(TA)}^2 / \bar{m} v_{LA(TA)}^2 \theta_{D,LA(TA)}$. Third, the scattering rate of phonons from isotopic mass-disorder which is given by[11] $\tau_{I,LA(TA)}^{-1}(\omega) = V g_2 \omega^4 / 4\pi v_{LA(TA)}^3$. Fourth, is the phonon-surface scattering which was first investigated by Casimir[18] and later extended by Berman *et. al.*[19] to include the effect of non-zero specularity factor (P). The phonon surface scattering rate can be expressed as[12–14] $\tau_{S,LA(TA)}^{-1} = v_{LA(TA)} / \mathcal{L}_{eff}$. $\mathcal{L}_{eff}$ is the effective width of the heat conductor.

**Generation mechanisms:** As mentioned in the main text, the earlier works on theoretical formalism of thermal transport in Si NWs argued that the decay of the optical phonons into acoustic phonons at an energy $\hbar\omega$ can be considered as a generation rate of acoustic phonons, which partially counteracts their scattering rate at the same energy[9]. Now, in majority of materials including Si, the anharmonic scattering and the isotope scattering of the optical phonons are linked to one another. It was argued that mass-disorder can open up additional scattering channels for the optical phonons, leading to an enhancement in their anharmonic scattering rate[20–22]. Therefore, while treating this mechanism in the isotopically mixed $^{28}Si_x\,^{30}Si_{1-x}$ NWs, a convolution of the two mechanisms (three-phonon anharmonic scattering and scattering from mass-disorder) needs



to be considered. The analytical expression for the generation rate of acoustic phonons (negative scattering rate of optical phonons) in presence of the convolution is given by[9] $\tau^{-1}_{G^M_{LA(TA)}}(\omega, T) = C^M_{LA(TA)} \omega^3 g_2 [\coth(\hbar\omega/2k_B T)]^{1/2}$ (M is the superscript denoting isotopically mixed NWs). In the isotopically pure $^{30}$Si NWs, devoid of any isotope scattering (so no convolution with isotope scattering), the generation rate of the LA(TA) phonons can be expressed as[9] $\tau^{-1}_{G^P_{LA(TA)}}(\omega, T) = C^P_{LA(TA)} \omega^5 [\coth(\hbar\omega/2k_B T)]$.

Finally, the total acoustic phonon scattering rate was calculated according to the Matthiessen rule as

$$\frac{1}{\tau_{j=LA(TA)}(\omega, T)} = \left[ \frac{1}{\tau_{N,j}(\omega, T)} + \frac{1}{\tau_{U,j}(\omega, T)} + \frac{1}{\tau_{I,j}(\omega)} + \frac{1}{\tau_{s,j}} \right] - \frac{1}{\tau_{G^{M(P)}_j}(\omega, T)} \quad (E3)$$



## 4) Simulated $\kappa_T$ vs. T data for bulk $^{Nat}Si$, bulk $^{30}Si$, and bulk $^{28}Si_{0.4}{}^{30}Si_{0.6}$

To simulate the $\kappa_T$ vs. T data for the bulk semiconductors, the upper limit of the integral in equation (3) of the main manuscript was set to $\omega_{D,LA}$ for the LA phonons. The result of the simulations is shown in Fig. S2. It shows that the bulk $^{30}Si$ sample reach $\kappa_{max}$ at a temperature $T_{max}^{bulk}$ of 26.5 K. $\kappa_{max}$ of $^{Nat}Si$ is reduced by a factor of ~7.9 relative to $^{30}Si$ at 26.5 K, and $\kappa_T$ of $^{Nat}Si$ is reduced by ~9.5 % relative to $^{30}Si$ at 300 K. The isotope effect on $\kappa_T$ of $^{Nat}Si$ vanishes at ~5.5 K (see pink dotted line). These values lie within ±5% of the experimentally measured values in literature[15,23]. Note that no theoretical or experimental investigation regarding the evolution of $\kappa_T$ of isotopically mixed bulk Si exists in the literature. The model has therefore been extended to display the temperature evolution of $\kappa_T$ of bulk $^{28}Si_{0.4}{}^{30}Si_{0.6}$ (same isotopic composition as the NWs used in this work), as shown using the solid red curve in Fig. S2. $\kappa_{max}$ of bulk $^{28}Si_{0.4}{}^{30}Si_{0.6}$ is reduced by a factor of ~45.8 relative to $^{30}Si$ and by a factor of ~5.8 relative to $^{Nat}Si$, at $T_{max}^{bulk}$ of 26.5 K. At 300 K, $\kappa_T$ of bulk $^{28}Si_{0.4}{}^{30}Si_{0.6}$ is reduced by ~32 % relative to $^{30}Si$. The isotope effect on $\kappa_T$ of bulk $^{28}Si_{0.4}{}^{30}Si_{0.6}$ vanishes at ~2.6 K (see orange dotted line).



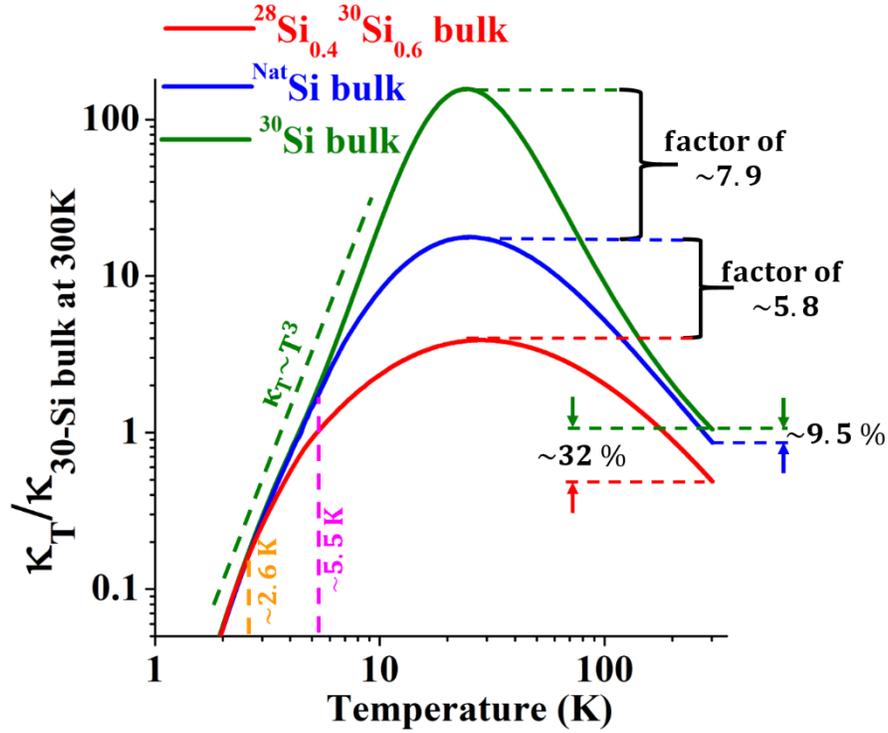

**Fig. S2 | Simulated $\kappa_T$ of bulk isotopically engineered Si**. Simulated $\kappa_T$ of bulk $^{30}$Si, and bulk $^{Nat}$Si, and bulk $^{28}$Si$_{0.4}$$^{30}$Si$_{0.6}$ as a function of ambient temperature, using equation (3) of the main manuscript. The temperature limit to the isotope induced phonon scattering is the temperature at which $\kappa_T$ of bulk $^{Nat}$Si, and bulk $^{28}$Si$_{0.4}$$^{30}$Si$_{0.6}$ merges with the $\kappa_T$ of bulk $^{30}$Si, and are shown using the pink and orange dotted lines, respectively. The relative changes in $\kappa_T$ between the bulk $^{30}$Si, bulk $^{Nat}$Si, and bulk $^{28}$Si$_{0.4}$$^{30}$Si$_{0.6}$ at 26.5 K and at 300 K are highlighted in the Figure.

## 5) Optimization of the cut-off frequency

As shown in Fig. S3 and discussed in the main manuscript, the quality of the fit to the experimentally obtained $\kappa_T$ vs. T data is rather poor when equation (3) of the main manuscript is integrated all the way up to the Debye frequencies of the modes. Note, during the fits, the parameters P and $C_{LA(TA)}^{M(P)}$ were used as fitting parameters and a bound of [0,1] was imposed on P. It is worth mentioning here that no amount of parameter adjustments led to any significant improvements in the quality of the fit when integrated up to $\omega_{D,LA(TA)}$. The reasons for are two fold: first, $M_j^{ph}(\omega)$ in equation (3) calculated using the Debye approximation can be shown to agree excellently with that calculated from the full phonon band-structure in most semiconductors, but only up to a certain energy, beyond which the approximation starts to divert[24]. Second, contribution that the Umklapp scattering makes to the total phonon lifetime in NWs. In bulk samples, the Umklapp scattering makes a dominant contribution to the total lifetime. The presence of phonon surface scattering in NWs at all temperatures means that the contribution of Umklapp scattering is relatively suppressed than in bulk samples. Consequently, considering a bulk-like phonon spectrum and integrating all the way up to the Debye frequency might overestimate the contribution of Umklapp scattering to the total phonon life-time in NWs[12]. The solution is therefore to chose a lower cut-off frequency of the modes which would get rid of the excess contribution of both $M_j^{ph}(\omega)$ as well as that of the Umklapp scattering to the total phonon life-time. The optimization was done on a trial and error basis by lowering the cut-off frequency of the modes by 1.0 THz. Fig. S3 shows a selected few examples, demonstrating the gradual improvement to quality of the fit as a result of the optimization. Also note that, that lowering the cut-off frequency of the TA mode leads to only a nominal change in the quality of the fit (see the pink and light green dot-dash-dot lines corresponding to $\omega_{D,LA} = 51.7$ THz, $\omega_{D,TA} = 26.4$ THz and $\omega_{D,LA} = 51.7$ THz, $\omega_{D,TA} = 21.4$ THz, respectively) that too in the high-temperature range ($> 180$ K). For this reason, during the cut-off optimization, the maximum frequency of the TA mode was maintained at $\omega_{D,TA} = 31.4$ THz, while only $\omega_{D,LA}$ was altered. This prevented having an additional unknown parameter (that is, $\omega_{C,TA}$) into the theoretical model. Note, every time the cut-of frequency was altered, the cut-off temperature, in the expression of $\tau_{U,LA(TA)}^{-1}(\omega, T)$ was also changed accordingly, from $\theta_{D,LA(TA)}$ to $\theta_{C,LA(TA)} = \hbar\omega_{C,LA(TA)}/k_B$.

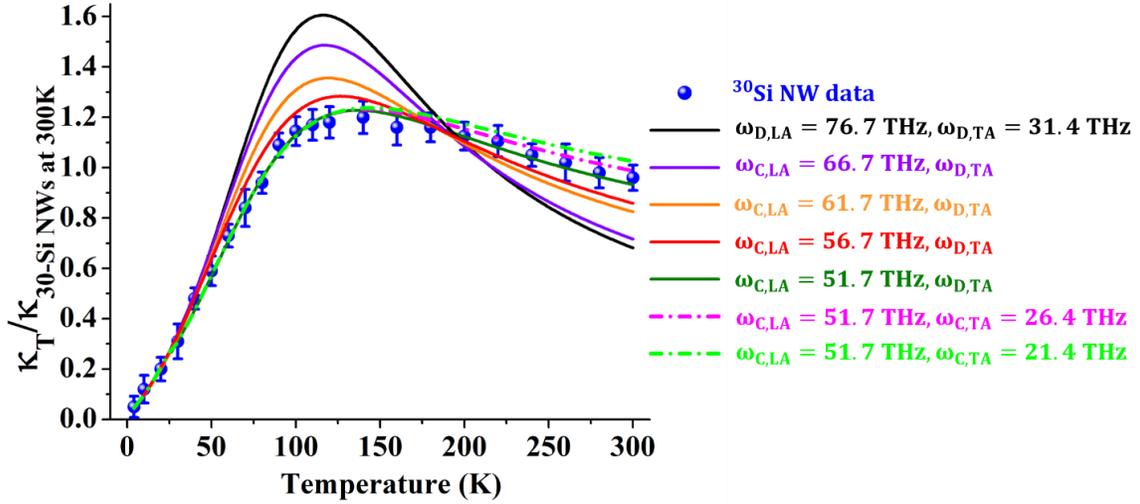

**Fig. S3│Cut-off optimization for the LDL model.** The cut-off optimization process with the experimentally obtained $\kappa_T$ *vs.* T data of the $^{30}$Si NWs. Selected few demonstrations are displayed using the colored lines (solid and dot-dashed), which are the best possible fits to the experimental data, obtained by choosing various values of $\omega_{D,LA}$ and $\omega_{D,TA}$ (as mentioned in the Fig. legend). The solid green line ($\omega_{D,LA} = 51.7$ THz, $\omega_{D,TA} = 31.4$ THz) is one depicted in Fig. 3(a) of the main manuscript.

**6) Simulated $\kappa_T$ *vs.* T data for NWs at intermediate values of $d_{NW}$**

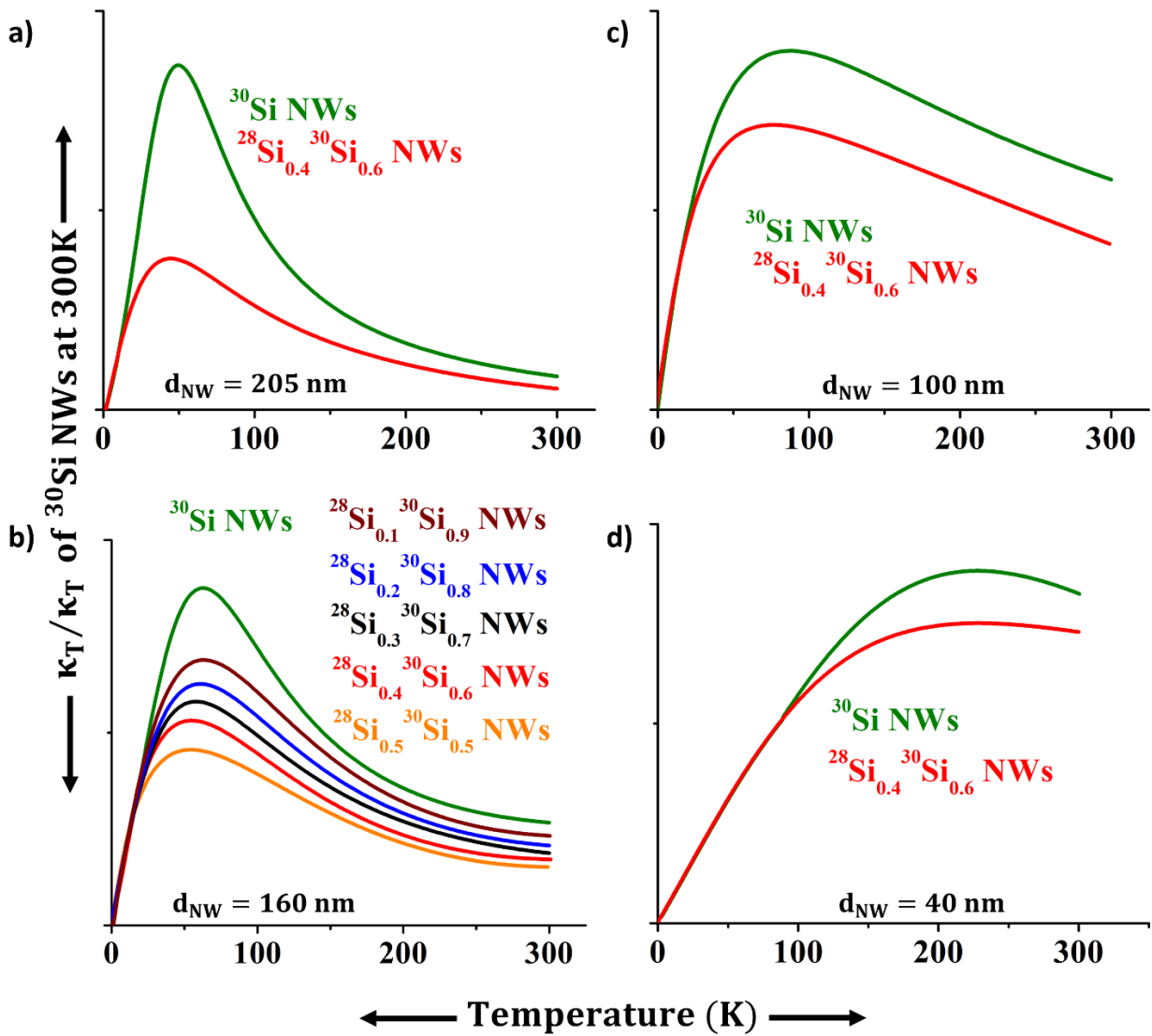

**Fig. S4 | $\kappa_T$ *vs.* T data for NWs at intermediate values of $d_{NW}$.** The simulated $\kappa_T$ *vs.* T data for isotopically pure and isotopically mixed NWs with diameters (a) $^{30}$Si NWs and $^{28}$Si$_{0.4}$$^{30}$Si$_{0.6}$ NWs with $d_{NW} = 205$ nm. (b) $^{30}$Si NWs and $^{28}$Si$_{0.1}$$^{30}$Si$_{0.9}$ NWs, $^{28}$Si$_{0.2}$$^{30}$Si$_{0.8}$ NWs, $^{28}$Si$_{0.3}$$^{30}$Si$_{0.7}$ NWs, $^{28}$Si$_{0.4}$$^{30}$Si$_{0.6}$ NWs, $^{28}$Si$_{0.5}$$^{30}$Si$_{0.5}$ NWs, with $d_{NW} = 160$ nm. (c) $^{30}$Si NWs and $^{28}$Si$_{0.4}$$^{30}$Si$_{0.6}$ NWs with $d_{NW} = 100$ nm. (d) $^{30}$Si NWs and $^{28}$Si$_{0.4}$$^{30}$Si$_{0.6}$ NWs with $d_{NW} = 40$ nm.

# 7) Variation of $R_{RT}^{d_{NW}}$, $T_{max}^{d_{NW}}$, and $R_{T_{max}}^{d_{NW}}$ as a function of $d_{NW}$

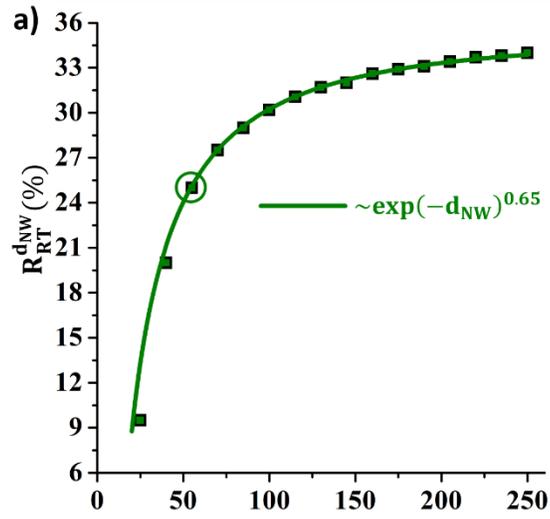

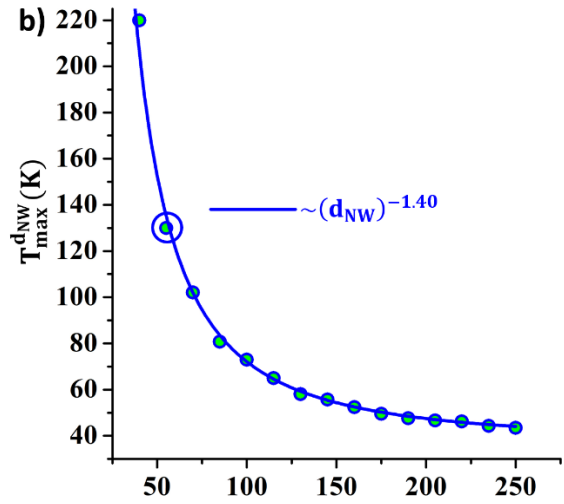

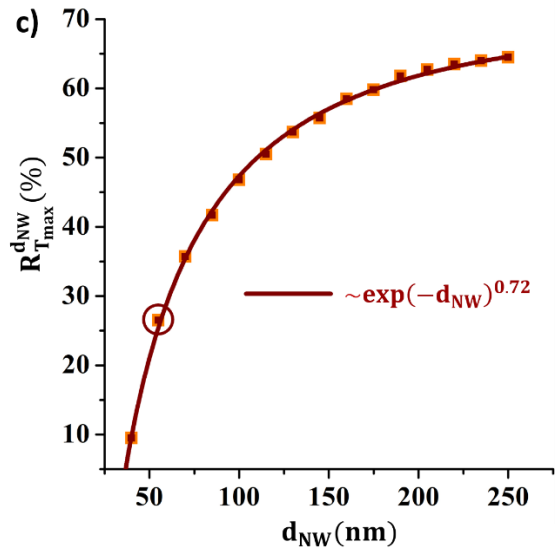

**Fig. S5 | Evoltuion of $R_{RT}^{d_{NW}}$, $T_L^{d_{NW}}$, and $T_{max}^{d_{NW}}$ as a function of the NW diameter.** (a) The evolution $R_{RT}^{d_{NW}}$ as a function of $d_{NW}$. The solid green line is an exponential fit to the simulated data, showing a $\sim\exp(-d_{NW})^{0.65}$ dependence on $d_{NW}$. (b) The evolution of $T_{max}^{d_{NW}}$ as a function of $d_{NW}$. The solid blue line is the fit to the simulated data, showing a $\sim(d_{NW})^{-1.4}$ dependence. (c) The evolution of $R_{T_{max}}^{d_{NW}}$ as a function of $d_{NW}$. The solid brown line is the fit to the simulated data, showing a $\sim\exp(-d_{NW})^{0.72}$ dependence. Note, that the NWs with $d_{NW}$ of 25 nm never reaches $T_{max}$ within the temperature range of 300 K considered in this work (see Fig. 3(b) of the main manuscript). Consequently, the data in (b) and (c) starts at $d_{NW} = 40$ nm and does not show the data point corresponding to $d_{NW} = 25$ nm. In (a)-(c), the simulated data point corresponding to $d_{NW} = 55$ nm (the average diameter of the NWs investigated experimentally in this work) is highlighted using the colored circles. Note that $T_{max}^{d_{NW}}$ of the $^{28}Si_x^{30}Si_{1-x}$ NWs and the $^{30}Si$ NWs are not exactly at the same point. $T_{max}^{d_{NW}}$ of the $^{28}Si_x^{30}Si_{1-x}$ NWs was always found to be at a slightly lower temperature (4–6 K) than that of the $^{30}Si$ NWs. $T_{max}^{d_{NW}}$ in (b) and $R_{T_{max}}^{d_{NW}}$ in (c) are relative to that of the isotopically pure $^{30}Si$ NWs.